\documentclass[a4paper,11pt]{article}
\usepackage{pos}

\title{Bridge++ 2.0: Benchmark results on supercomputer Fugaku}

\author[a]{Tatsumi Aoyama}
\author*[b]{Issaku Kanamori}
\author[c]{Kazuyuki Kanaya}
\author[d,e]{Hideo Matsufuru}
\author[f,g]{Yusuke Namekawa}

\affiliation[a]{%
Institute for Solid State Physics, University of Tokyo,
Chiba 277-8581, Japan}

\affiliation[b]{%
 RIKEN Center for Computational Science (R-CCS), \\
 7-1-26, Minatojima Minamimachi, Kobe 650-0047, Japan}

\affiliation[c]{Tomonaga Center for the History of the Uiverse, University of Tsukuba, Ibaraki 303-8571, Japan}

\affiliation[d]{Computing Research Center, High Energy Accelerator Research Organization (KEK),\\
 1-1 Oho, Tsukuba 305-0801, Japan}

\affiliation[e]{School of High Energy Accelerator Science, Graduate University for Advanced Studied (SOKENDAI),\\
 1-1 Oho, Tsukuba 305-0801, Japan}

\affiliation[f]{Education and Research Center for Artificial Intelligence and Data Innovation,
Hiroshima University, Hiroshima 730-0053, Japan}

\affiliation[g]{Department of Physics, Faculty of Science, Kyoto University, Kyoto 606-8502, Japan}

\emailAdd{kanamori-i@riken.jp}

\abstract{%
Bridge++ is a general-purpose code set for lattice QCD simulations
aiming at a readable, extensible, and portable code while
keeping practical high performance.
The new version 2.0 employs machine-dependent optimization,
enabling flexible data layout in float/double precision,
while it was fixed layout and only with the double precision in previous versions.
We report the performance on supercomputer Fugaku
with Arm A64FX-SVE architecture by Fujitsu.

}

\FullConference{%
The 39th International Symposium on Lattice Field Theory,\\
8th-13th August, 2022,\\
Rheinische Friedrich-Wilhelms-Universität Bonn, Bonn, Germany
}

\begin{document}
\maketitle

\section{Introduction}

Bridge++ \cite{Bridge,Ueda:2014rya,Ueda:2014zsa} is a C++ code set for simulations of lattice gauge theory.
It is based on the object oriented design and intended to be readable, extensible,
and portable while keeping reasonable practical performance.
After the releas of version 1.0 in 2009, a lot of new architectures have appeared
whose performance can be brought out only with specific optimization techniques.
We have therefore decided to provide a major update of the code set
that contains optimizations to new machines in the forthcoming version 2.0 \cite{Akahoshi:2021gvk}.
The original fixed data layout with double precision floating point numbers
is generalized to flexible data layouts in double or single precisions.
Exploratory implementations and tuning have been studied for
the SIMD architecture with Intel AVX-512 \cite{Kanamori:2017tlp,Kanamori:2017urm,Kanamori:2018hwh_inbook},
GPU architectures with OpenACC \cite{Motoki:2014a, Motoki:2016rii} or OpenCL \cite{Ueda:2015mda, MOTOKI2014170}, the PEZY-SC
many-core accelerator \cite{Aoyama:2016a},
and a vector architecture of NEC SX-Aurora TSUBASA \cite{Akahoshi:2021gvk}.
In the framework of Bridge++ version 2, these machine-specific implementations
are provided as alternative codes to the original implementation 
which takes over the specific tasks such as calculation of fermion propagators.

One of our main targets of this update is another SIMD-type architecture,
Arm A64FX-SVE by Fujitsu \cite{a64fx}.
In this article, we focus on our tuning and benchmark results on the
supercomputer ``Fugaku'' installed in RIKEN Center for Computational
Science, which is the leading system employing this architecture.
Fugaku is a massive parallel system composed of 158,976 A64FX processors
on a six-dimensional mesh-torus network.
Each processor consists of 48 computational cores accompanied by 2 or 4
assistant cores, an on-chip 32 GB HBM memory, and the network interface
(TofuD interconnect) also assembled on the chip.
These cores are grouped into 4 Core Memory Groups (CMGs) each having
12 computational cores and 8 GB memory.
MPI parallelization in units of this CMG is assumed to achieve the best
performance.
Each computational core has 32 SIMD registers and two sets of SIMD
floating-point operation units of 512-bit length, for which the scalable
vector extension (SVE) of the Arm instruction set architecture provides
efficient arithmetic execution.
The theoretical peak performance of one processor results in
3,072 GFlops for double precision at the normal mode with 2.0 GHz,
which corresponds to 488 PFlops for the total Fugaku system.

To maximally make use of the hardware potential of the Fugaku supercomputer,
so-called co-design development has been adopted that develops hardware and
software concurrently against several target applications \cite{9355239}.
Lattice QCD simulation is one of the target applications and the co-design activity accomplished development of the QCD Wide SIMD library (QWS) \cite{Ishikawa:2021iqw}.
The QWS library implements a domain-decompose linear equation solver
for the clover fermion and achieved sustained performance of more than 100 PFlops.
The domain-decomposed preconditioner in QWS was exploited in a multi-grid solver
for Fugaku \cite{Ishikawa:2021day}.
Other QCD codes have been also tuned for A64FX, as exemplified by the Grid library
\cite{Meyer:2021uoj}.
The status of lattice QCD code for Fugaku and ARM architecture is reviewed in
\cite{nakamuraLat2021}.

\section{Implementation for Fugaku}

While the QWS library achieves good performance, its application is restricted:
it implements only the clover fermion operator in a linear equation solver,
adopts one-dimensional SIMD packing that strongly restricts the volume setup
on each node, and achieves the best performance only with a specific choice of
volume parameters.
Since Bridge++ is a general purpose code set, we intend to relax these
restrictions while being guided by the QWS implementation to achieve
sufficient performance.

As for the fermion operators, in addition to the Wilson/clover action,
we implement the standard staggered and M\"obius domain-wall actions.
For each of these fermion operators, we also implement their even-odd versions,
$D_{\mathrm{ee}}$, $D_{\mathrm{oo}}$, $D_{\mathrm{eo}}$, and $D_{\mathrm
{oe}}$ in the following decomposition.
\begin{align}
D^{\text{(even-odd)}}
&=1- D_{\mathrm{ee}}^{-1} D_{\mathrm{eo}}   D_{\mathrm{oo}}^{-1} D_{\mathrm{oe}},
&
\text{with full operator } 
D&=
\begin{pmatrix}
 D_{\mathrm{ee}} &   D_{\mathrm{eo}} \\ 
 D_{\mathrm{oe}} &   D_{\mathrm{oo}}
\end{pmatrix}.
\label{eq:even-odd}
\end{align}
$D^{\text{(even-odd)}}$ is used in the even-odd preconditioned solver that achieves
better efficiency in most cases.
While one can construct the domain-wall fermion operator employing the Wilson
operator, we additionally implement a code that treats the fifth-dimensional
coordinate as an inner degree of freedom.
The latter implementation achieves better performance than the former generic
construction.

\begin{figure}
\centering 
\includegraphics[width=0.4\linewidth]{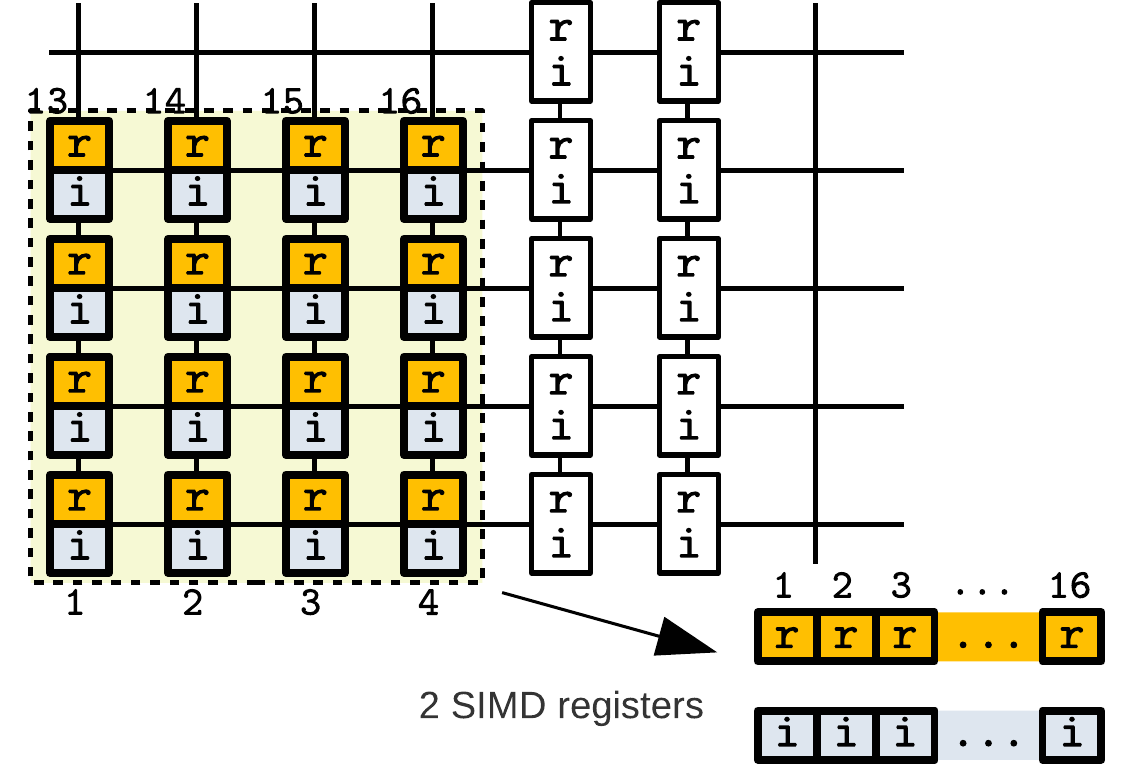}
\caption{
Data layout.  The site degrees of freedom are packed to the vector register
\cite{Akahoshi:2021gvk}.
Two-dimensional tiling is used (the above is $4\times 4$ for single precision case).
We use separate registers to real and imaginary parts of complex numbers. 
}
\label{fig:layout_QXS}
\end{figure}

Following the implementation in QWS, we distribute the real and imaginary
parts of a complex number in separate SIMD registers.
This differs from the implementation for AVX-512 instruction set architecture
which possesses proper instructions for complex numbers inside SIMD vectors
\cite{Kanamori:2018hwh_inbook}.
The SIMD parallelization is applied to the lattice site degrees of freedom,
for which we adopt two-dimensional tiling in $x$- and $y$-directions
as displayed in Fig.~\ref{fig:layout_QXS}.
This is in contrast to the case of QWS which packs the sites in the
$x$-direction into each register, and thus the local extent in $x$-direction
must be a multiple of 16 (or 32 for the even-odd domain decomposition to work).
Two-dimensional tiling relaxes this constraint, which is particularly convenient
for the multi-grid solver that requires an operator on the coarse-grained
lattice.
While the SVE does not assume a fixed SIMD length, our implementation
explicitly assumes the 512-bit length that allows $16\times 1$, $16\times 1$, 
$8\times 2$, $4\times 4$, and $2\times 8$ $(x,y)$-tiling in the case of
single precision.
Apart from the 2D tiling, the data layout is in the same manner as the QWS
implementation including the convention of $\gamma$ matrices (except for
$\gamma_5$ with a negative sign), which requires data conversion before and
after calling the codes for A64FX.

After implementing the operations in units of the SIMD vector, which is
represented as \texttt{struct} objects, we apply the Arm C-Language Extension (ACLE)
to enforce the SIMD execution.
In particular masked load to SIMD register making use of the predicate mechanism
is useful to implement the shift of field in the $x$- and $y$-directions.
We also apply manual prefetching to the even-odd domain-wall operator.
Application of the prefetch to the other operators is also planned.

The communication between neighboring nodes is implemented by using MPI
persistent communication.
Alternatively to the standard MPI library, we make use of the Fujitsu extension
of the MPI persistent communication, which accelerates the overlap of
communication and computation by using the assistant cores.
The thread parallelization model for MPI communication is
\verb|MPI_THREAD_FUNNELED| so that the only master thread calls
the MPI functions.
Assuming that the assistant core absorbs communication overhead,
the tasks for the site loop are uniformly assigned to the threads
inside each MPI process.

In addition to the standard fermion operators,
we show a benchmark result for the multi-grid solver for the
clover fermion \cite{Kanamori:2021rwy}.
We adopt a variant of the DD$\alpha$-AMG algorithm
\cite{Frommer:2013fsa} and currently implement a two-level multi-grid.
The outer solver is Flexible BiCGStab (FBiCGStab) for which 
single precision multi-grid preconditioner is applied.
The coarse grid solver is BiCGStab and the smoother is
Schwartz Alternating Procedure (SAP).
Both the coarse and fine lattice operators adopt the same SIMD data
layout.
For Fugaku, we adopt the SAP preconditioner in QWS as 
a smoother \cite{Ishikawa:2021day}.
The Jacobi method with fixed numbers of iteration is used to solve
the subsystems in the SAP.
The block size for SAP in QWS is half of the local volume
so that every process has only two blocks.
This is different from the original DD$\alpha$-AMG, where each block
corresponds to one coarse lattice site and each process has
typically O(100) blocks or more.
If the local volume does not allow to use QWS, 
we can use Bridge++ implementation of SAP preconditioner.
The block size of the SAP in Bridge++ is similar to DD$\alpha$-AMG
and a fixed iteration MINRES solver is used to solve the subsystem.

\section{Benchmark Results}

The benchmark performance is measured on Fugaku with the normal mode
(2.0 GHz)\footnote{
The frequency of the I/O nodes is 2.2 GHz.},
unless otherwise stated.
We use the Fujitsu compiler on the language environment version 1.2.35,
in the Clang-mode with the optimization level \texttt{-Ofast}.
To reduce the effect of the shape of MPI processes in observing 
the scaling, all the neighboring communications in eight directions are
always performed, even if only one MPI process exists in that direction.
Four MPI processes are assigned to each node ({\it i.e.}, one process
per one CMG).
This way of assignment is usually faster than using 1 MPI process per node (in fact it is for Bridge++), since memory access over the CMGs is slower than accessing the memory
on the same CMGs.
The MPI rank map guarantees that the logical neighbors on 4-dimensional
torus correspond in the physically neighboring node or on the same node
so that the neighboring communication is always within a single hop.
As each CMG has 12 computation cores and no hyper-threading is supported,
we use 12 OpenMP threads in each MPI process.

\begin{figure}
\vspace*{-1mm}

\center
\includegraphics[width=0.44\linewidth]{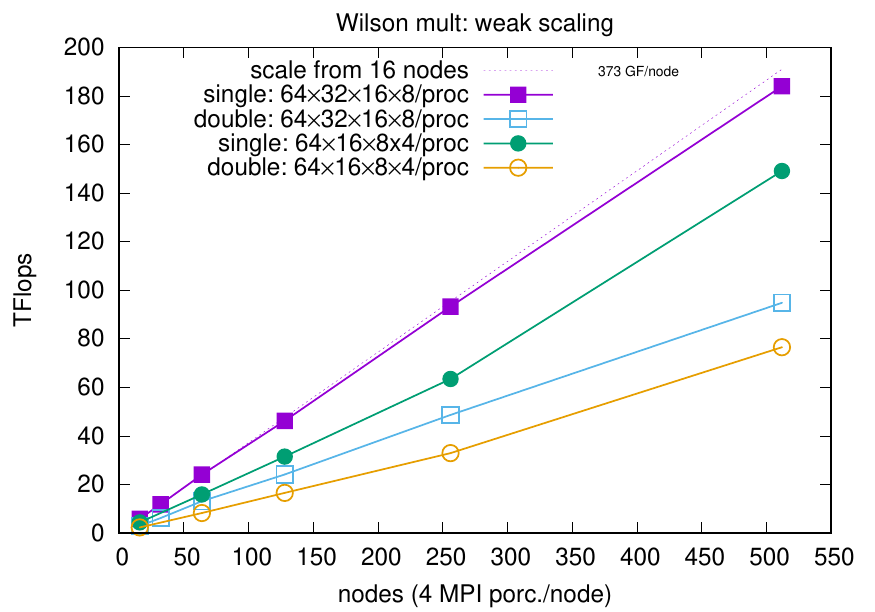}
\hfil
\includegraphics[width=0.44\linewidth]{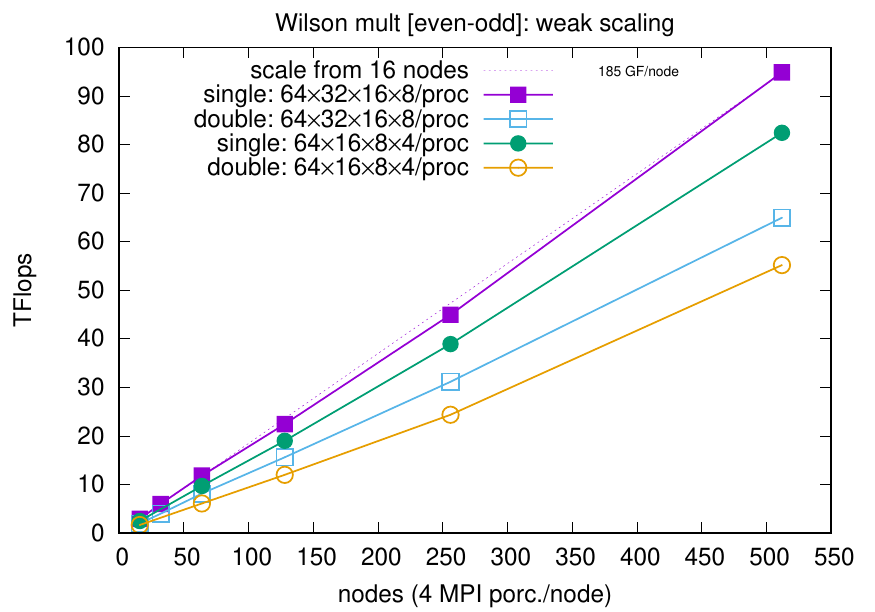}
\vspace*{-2mm}

\center
\includegraphics[width=0.44\linewidth]{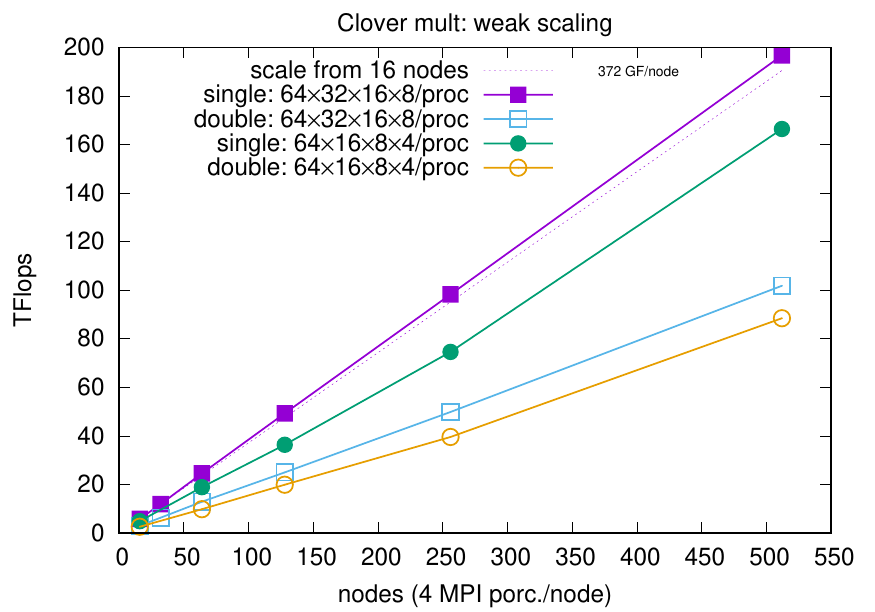}
\hfil
\includegraphics[width=0.44\linewidth]{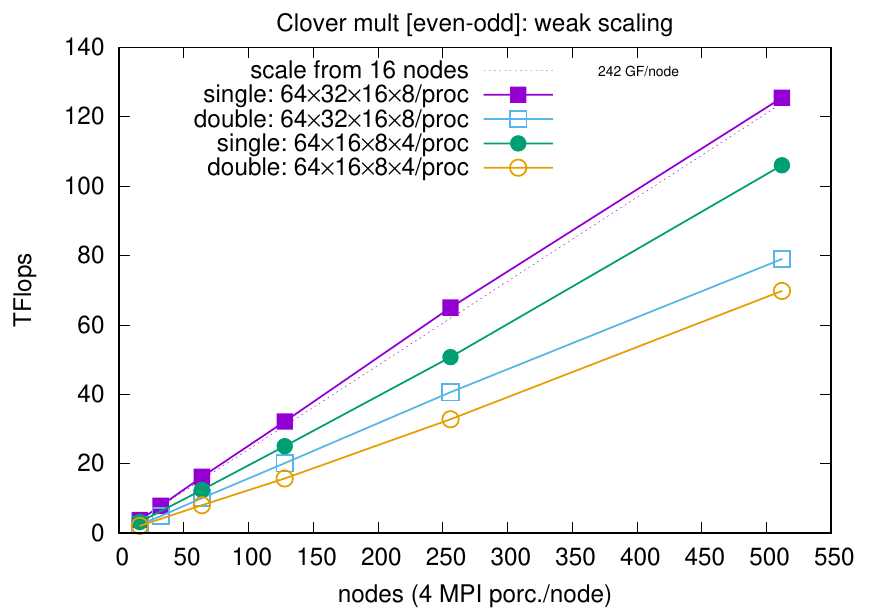}
\vspace*{-2mm}

\center
\includegraphics[width=0.44\linewidth]{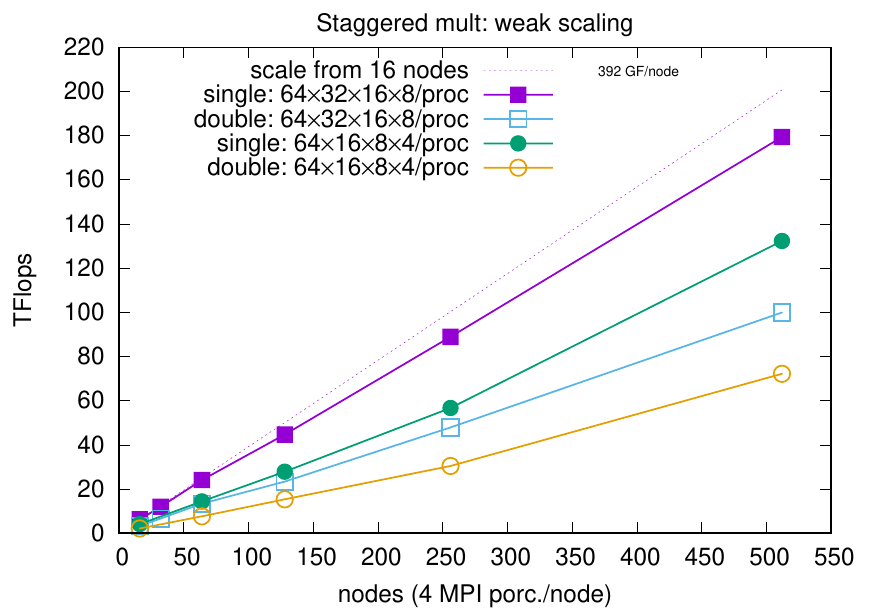}
\hfil
\includegraphics[width=0.44\linewidth]{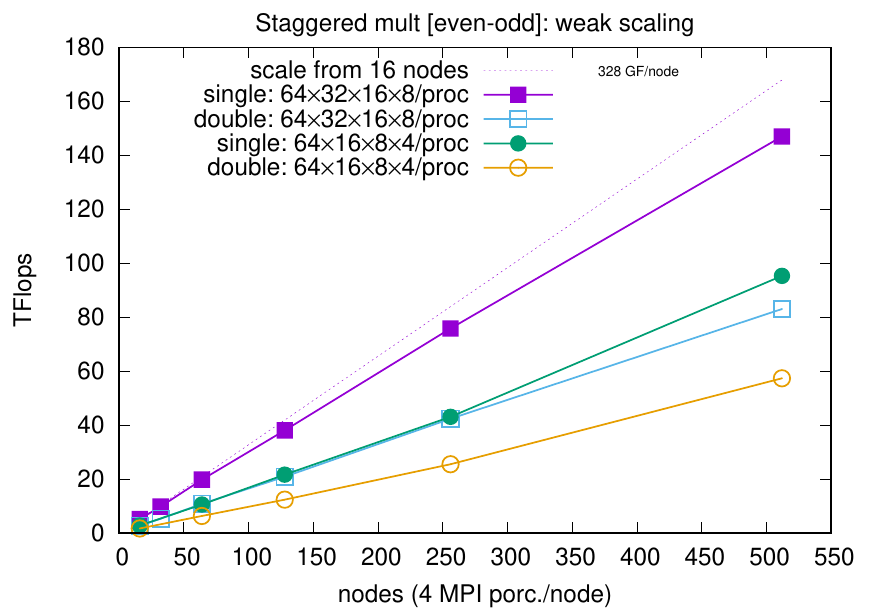}
\vspace*{-2mm}

\center
\includegraphics[width=0.44\linewidth]{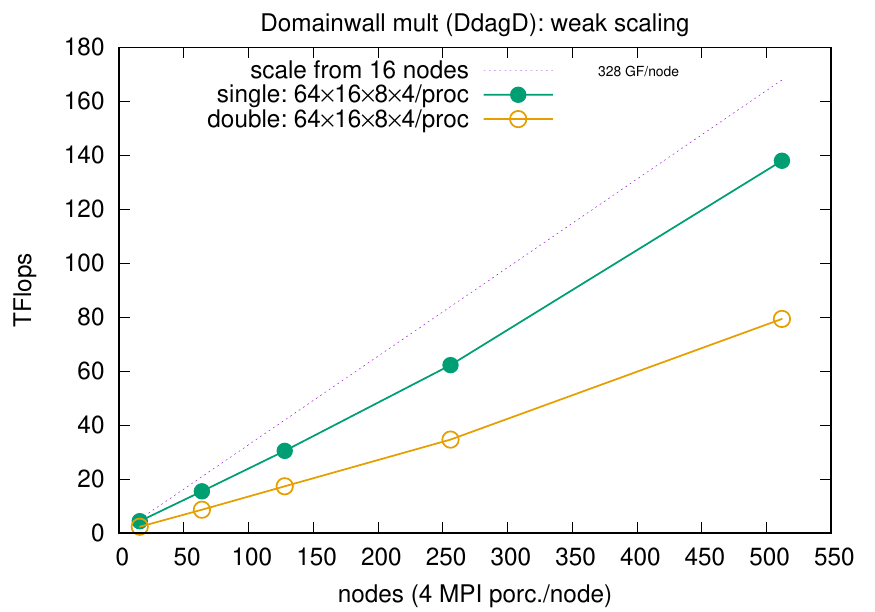}
\hfil
\includegraphics[width=0.44\linewidth]{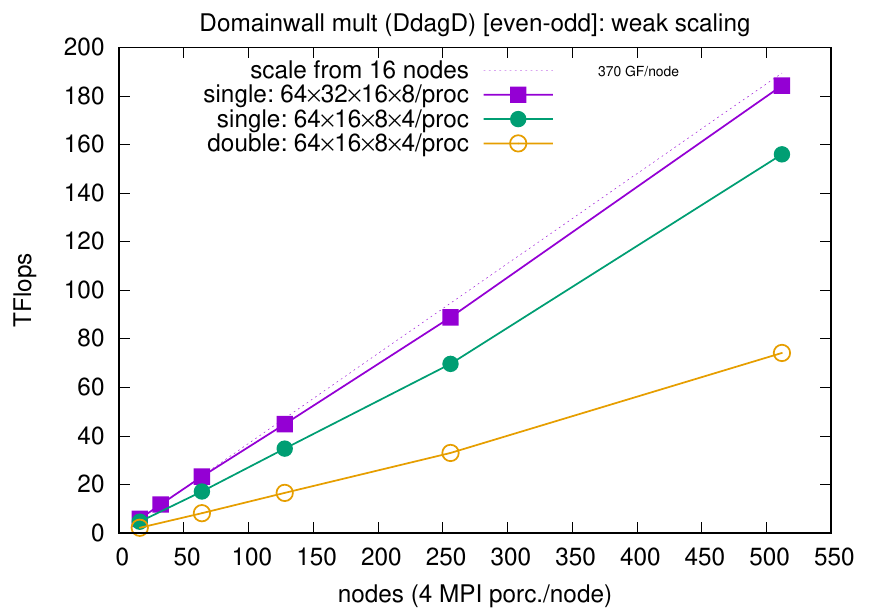}

\caption{Weak scaling of Dirac operator multiplications, from 16 nodes to 512 nodes. From the top panels to the bottom panels: Wilson, Clover, Staggered and Domainwall operators.  The left panels are full operators and right panels are even-odd preconditioned operators.
The lines are guide for the eye.  Each plot has two local volumes, $64\times 32 \times 16 \times 8$ (square) and $64\times 16 \times 8 \times 4$ (circle),
and both single (filled symbols) and double (open symbols) precisions.
The double precision Domainwall operators with larger local volume are missing 
due to the memory size limitation.
The 5th extension of the Domainwall operator is $L_s=8$.  }
\label{fig:weak_scaling_mult}
\end{figure}

Figure \ref{fig:weak_scaling_mult} shows the weak scaling behavior of
the performance for multiplication of various Dirac operators
up to 512 nodes.
In the right column, results for the even-odd operators $D^{\text{(even-odd)}}$
in Eq.~(\ref{eq:even-odd}) are displayed.
We measure the sustained performance in single and double precisions
for two local volumes on each node:
$64\times 32 \times 16 \times 8$ and $64\times 16 \times 8 \times 4$.
The extension in the fifth direction of the domain-wall operator
is set to $L_s=8$.
For the domain-wall fermion in double precision, the results for
the larger local volume are missing due to the memory size limitation.
All the operators exhibit good scaling behavior.
The results for the larger local volume (square symbols in the plot)
give better performance than those for the smaller volume (circles).
The single precision performance of the standard operators for the larger
local volume is around 300 GFlops/node (domain-wall) to 380 GFlops/node
(clover) on 512 nodes.\\
We notice that except for the domainwall operators the even-odd
preconditioned operators exhibit significantly lower
performance%
\footnote{
This performance drop of the even-odd operator has disappeared after
the post-conference tuning \cite{10.1145/3581576.3581610}.}.
This is mainly because of insufficient SIMD tuning in these operators,
which caused unnecessary gather and scatter operations.

\begin{figure}
\center
\includegraphics[width=0.43\linewidth]{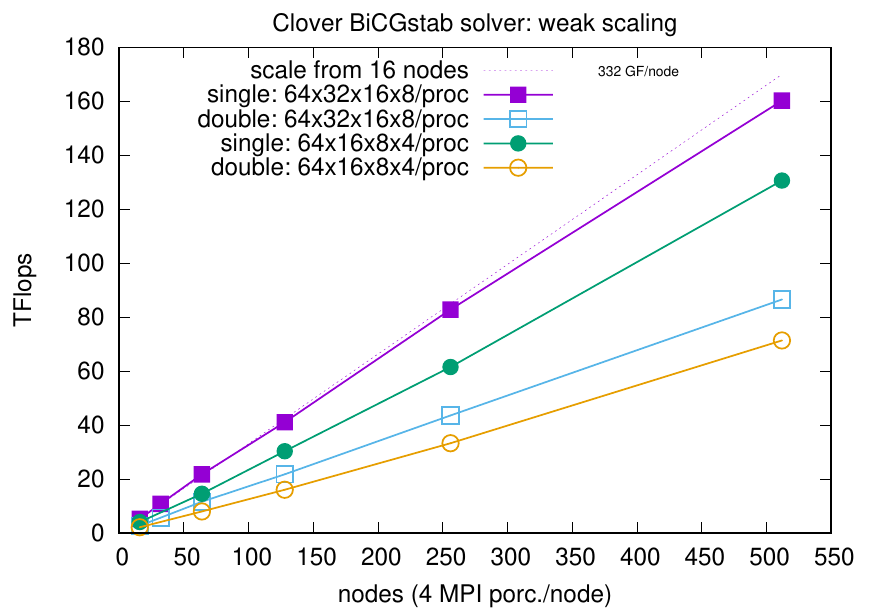}
\hfil
\includegraphics[width=0.43\linewidth]{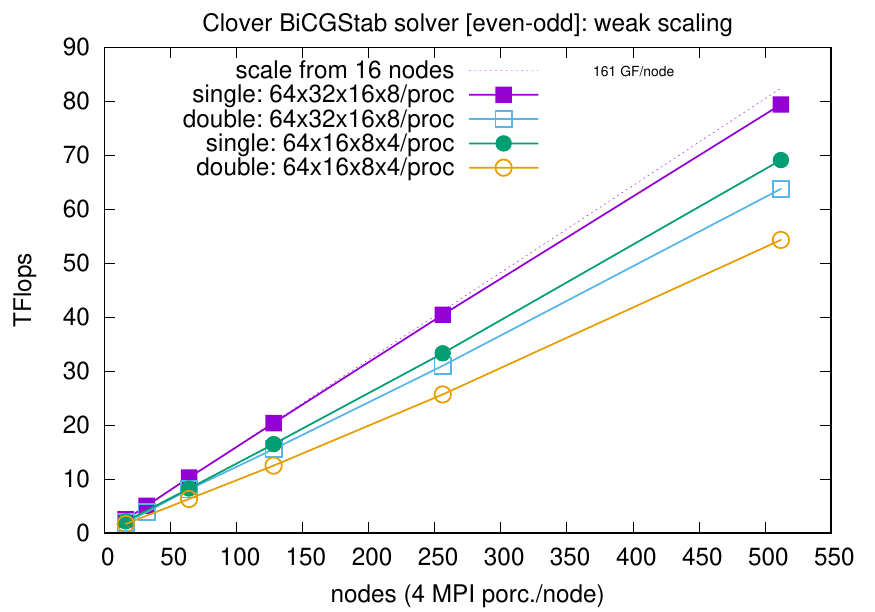}

\center
\includegraphics[width=0.43\linewidth]{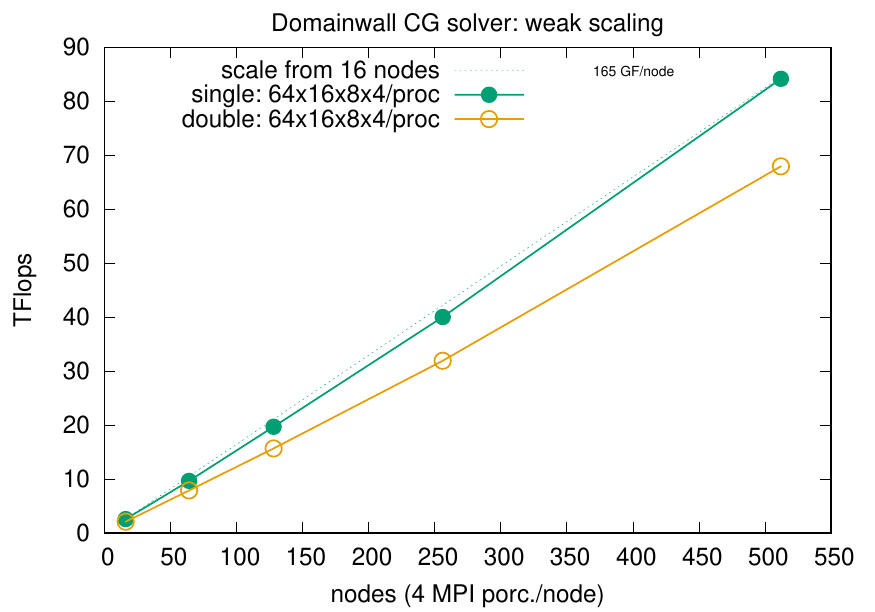}
\hfil
\includegraphics[width=0.43\linewidth]{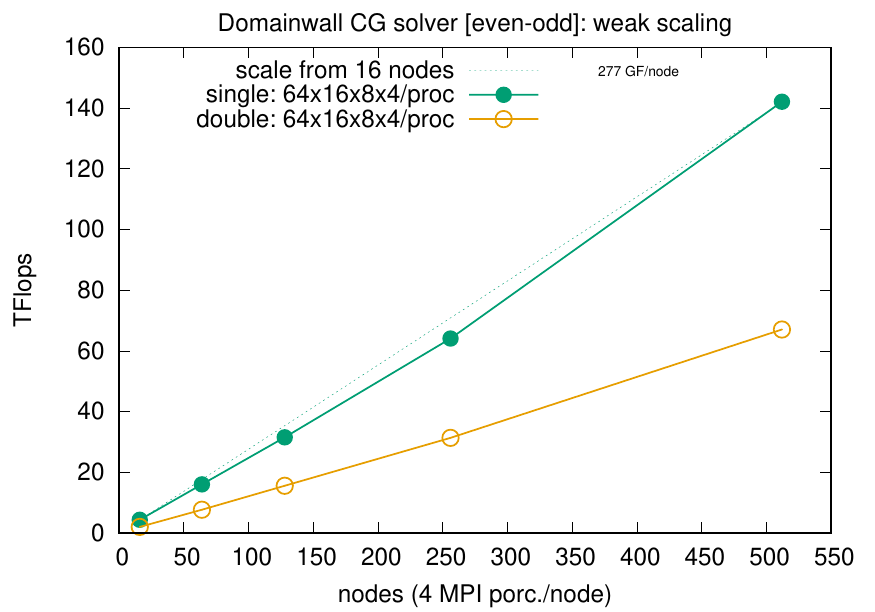}

\caption{Weak scaling of solver.
  The upper panels are BiCGStab solver for clover fermion and the lower panels are CG solver for Domainwall solver.  The details are the same as Fig.~\ref{fig:weak_scaling_mult}.}
\label{fig:weak_scaling_solve}
\end{figure}

Figure \ref{fig:weak_scaling_solve} shows the sustained performance
for linear equation solvers of the clover (BiCGStab algorithm)
and the domain-wall (CG) fermions.
The observed weak scaling behaviors are almost perfect, showing MPI
rank map works efficiently.
Compared to the Dirac operator multiplications, the performance of the
solver is lower in GFlops.
It is because the linear algebra in the solver requires
more byte-per-flop (B/F) ratio and global reductions.
In the case of Clover fermion, for example, 
the solver performance is about 65\% of 
the Dirac operator multiplication in single precision.
Note that the performance on Fugaku is much better than that on
Intel Xeon Phi Clusters.  The permanence of the solver
measured on Oakforest-PACS,
a massive parallel machine with Intel Xeon Phi Knights Landing (KNL), was
about or less than 50\% of the Dirac operator multiplications: 45\% 
in the case of single precision clover operator on 32 nodes \cite{Kanamori:2018hwh_inbook}.
The better performance on Fugaku is due to the higher memory bandwidth
 (B/F ratio 0.17 for Fugaku,  $\sim 0.08$ for KNL) and faster MPI reductions.
Global reductions on Fugaku with Tofu barrier execute reduction operations up to 3 elements simultaneously with the barrier synchronization, which is in fact efficient 
in the iterative solvers \cite{Ishikawa:2021iqw}.

\begin{figure}
\center
\includegraphics[width=0.43\linewidth]{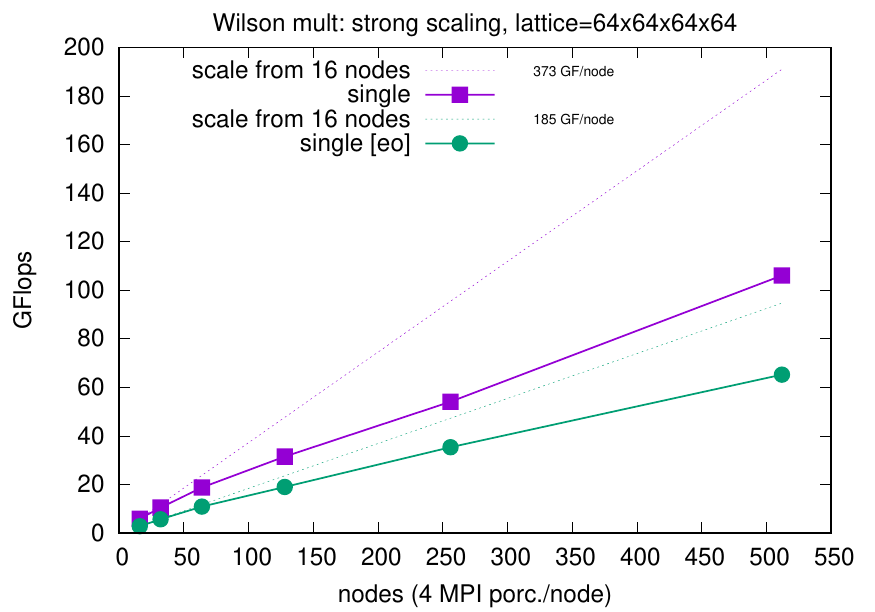}
\hfil
\includegraphics[width=0.43\linewidth]{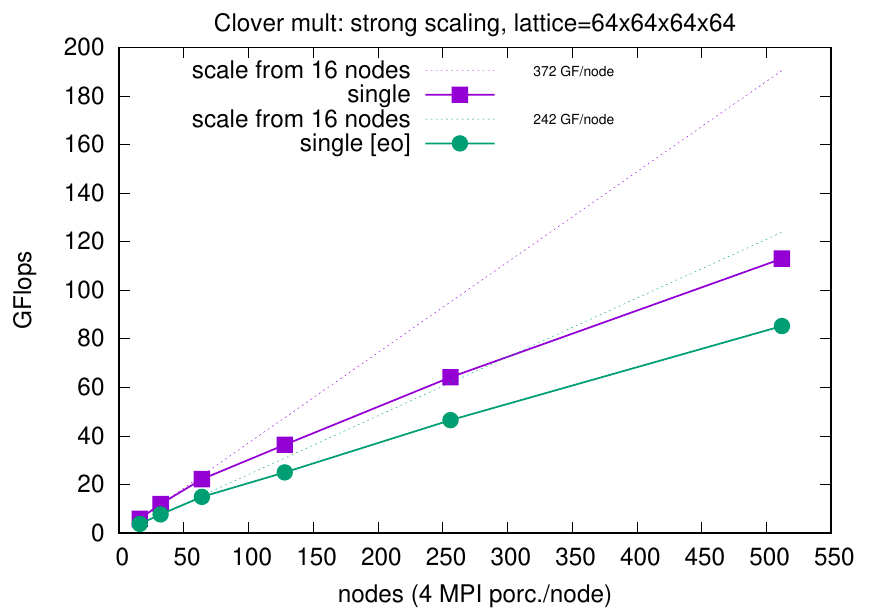}

\center
\includegraphics[width=0.43\linewidth]{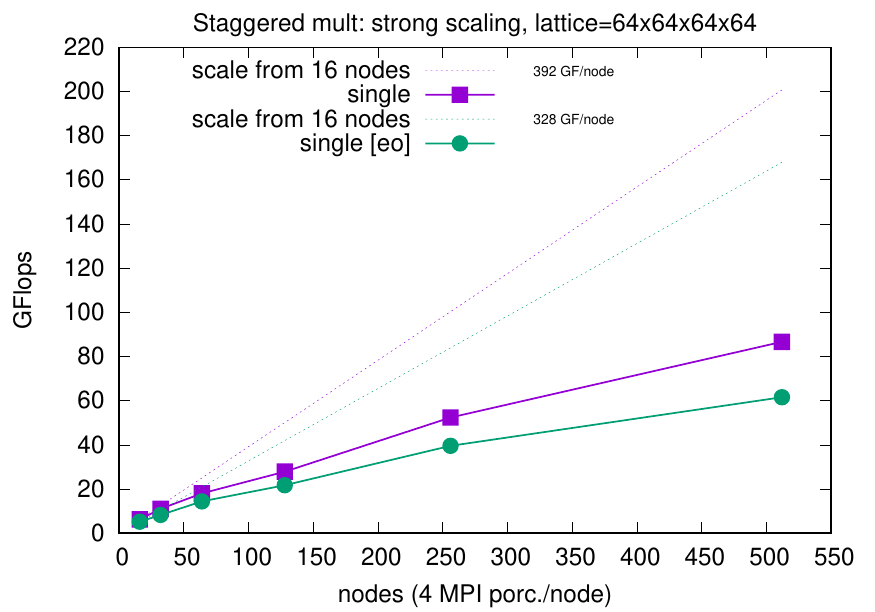}
\hfil
\includegraphics[width=0.43\linewidth]{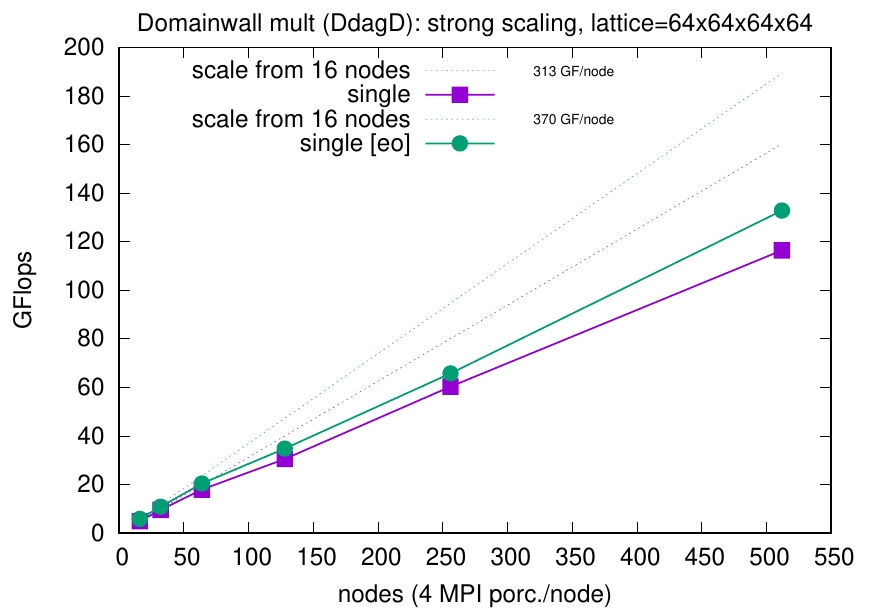}

\caption{
Strong scaling of Dirac operator multiplication.
From upper left to lower right: Wilson, clover, staggered and domainwall operator.
Dotted lines are ideal scaling from 16 nodes for single precision operators.
}
\label{fig:strong_scaling_mult}
\end{figure}

In Figure~\ref{fig:strong_scaling_mult}, the strong scaling plots are
displayed for the single precision Dirac operator multiplications
for a $64^4$ lattice on up to 512 nodes.
For this lattice size, we observe good scaling up to 64 nodes,
followed by acceptable scaling behavior up to the largest number of nodes
we tested without clear saturation.
This implies that the communication overhead is still not dominant
for this setup.

\begin{figure}
\center
\includegraphics[width=0.43\linewidth]{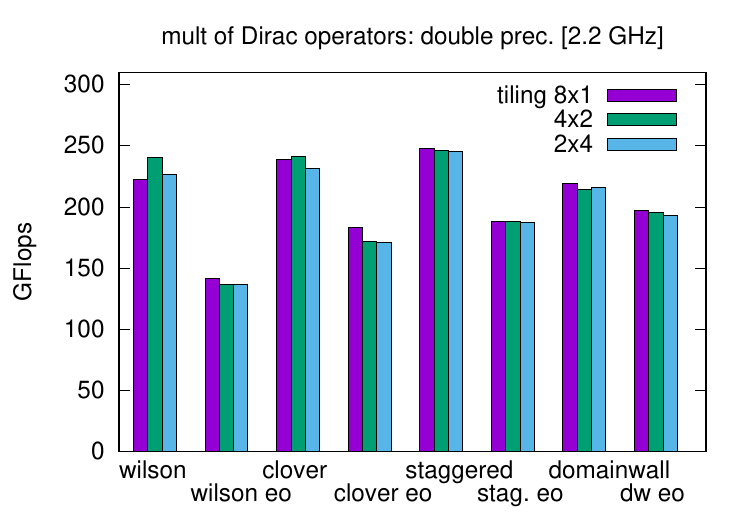}
\hfil
\includegraphics[width=0.43\linewidth]{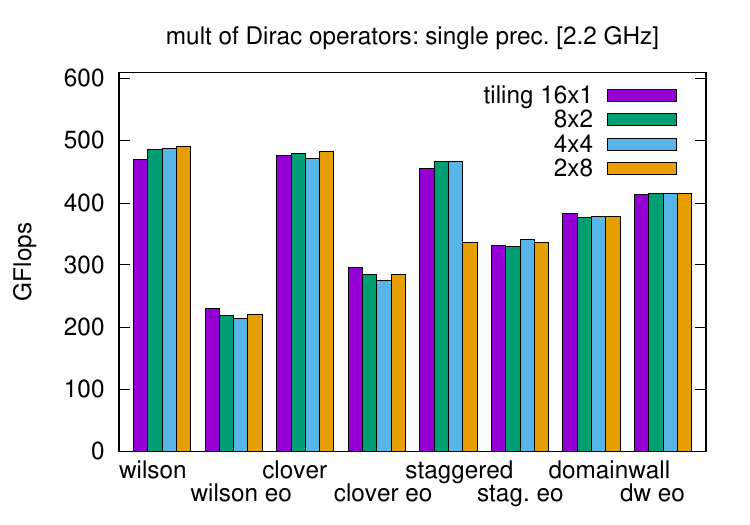}

\caption{
Performance of the Dirac operator multiplication with various SIMD tilings on single node. The lattice volume is $64\times 16 \times 16 \times 8$. The MPI process size is $1\times 1\times 2\times 2$ and  the irrelevant communication
in $x$- and $y$-directions are switched off.  The left panel is double precision and the right panel is single precision.  The data in these two plots were taken at `Flow' in Nagoya University,
of which 2.2 GHz frequency is higher than the normal mode of Fugaku (2.0 GHz).
}
\label{fig:mult_tiling_dep}
\end{figure}

The dependence on the details of two-dimensional SIMD tiling is
investigated in Fig. \ref{fig:mult_tiling_dep}.
The figure shows that there is no significant difference among
different tiling shapes.
The redundant neighboring communications are switched off in this figure.
This result indicates that one can flexibly choose the tiling and
local volume without the loss of performance.

\begin{figure}
\center
\includegraphics[width=0.43\linewidth]{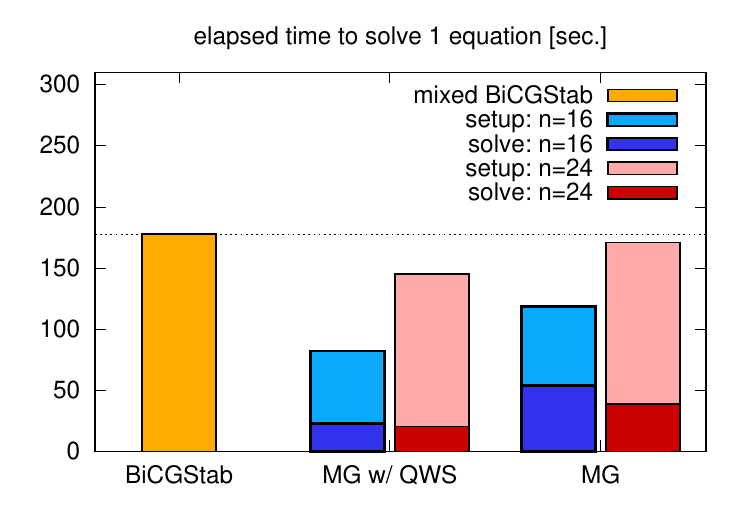}
\hfil
\includegraphics[width=0.43\linewidth]{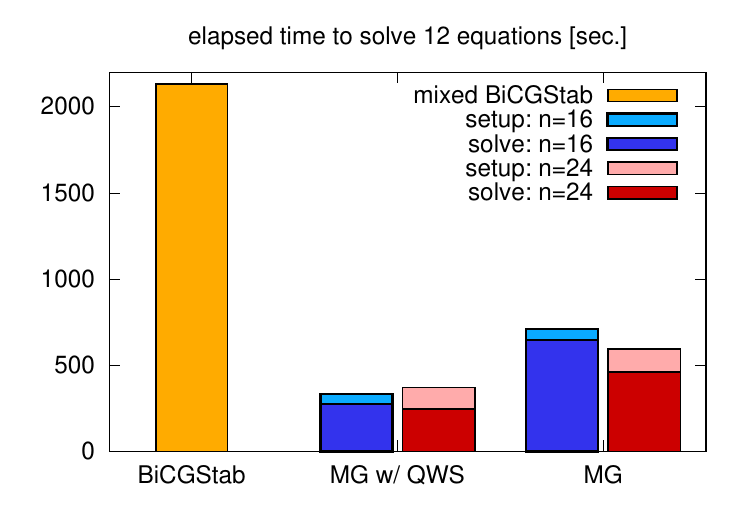}

\caption{
Performance of the multi-grid solver, 
on $96^4$ lattice with $M_\pi=145$ MeV configuration from \cite{Ishikawa:2015rho}.
The left panel is the elapsed time to solve 1 equation and 
the right panel is to solve 12 equations, including setup time denoted light colors, with 16 or null space vectors (denoted by $n$) to build the coarse system.
For comparison, the timing with mixed precision BiCGStab solver is plotted.
}
\label{fig:MGsolver}

\end{figure}

The performance of the multi-grid solver together with mixed BiCGStab solver
is plotted in Fig.~\ref{fig:MGsolver}.
The configuration is from PACS collaboration \cite{Ishikawa:2015rho},
of which lattice size is $96^4$ and the pion mass is $M_\pi=145$ MeV.
We tested both implementations with and without QWS, and
two different numbers of null space vectors (denoted $n$ in the plots) to build the coarse system.
Using QWS makes solving time significantly faster, it accelerates by a factor $\sim 2$--$2.5$.
Within this comparison, even solely solving one equation with multi-gird solver 
is faster than mixed BiCGStab including the large setup overhead to build the coarse system.
The setup overhead becomes irrelevant in solving O(10) or more equations with the same
background gauge field, as shown in the right panel of the figure.
Note that the full solver from QWS can be faster than mixed BiCGStab solver in Bridge++.
The reimplementation of QWS for domain decomposed HMC solves the same equation 
in 52 sec. \cite{Ishikawa:2021day}, which is in fact faster than the time to solve 1 equation
using multi-grid solver including the setup costs, 82 sec. ($n=16$) or 145 sec. ($n=24$).
To solve 12 equations, however, the multi-grid solver with QWS takes only 330 sec. ($n=16$),
while the (reimplementation of) QWS takes more than 600 seconds.

\section{Summary}

The new version of Bridge++ has several extended data layouts, with which
the architecture specific implementation is made.
The implementation to A64FX, more specifically, 
to supercomputer ``Fugaku'', shows very good weak scaling in 
various Dirac operator multiplications and iterative linear solvers up to more than 500 nodes.
The multi-grid solver utilizes a preconditioner in QWS, which was
developed in the co-design activity for Fugaku, and achieves a better 
throughput than QWS for almost physical point configuration.
The code will be publicly available soon at the Bridge++ website \cite{Bridge}.

\subsection*{Acknowledgments}

This work is supported by JSPS KAKENHI (JP20K03961, JP21K03553), the MEXT as
`Program for Promoting Researches on the Supercomputer Fugaku' (Simulation for basic science:
from fundamental laws of particles to creation of nuclei) and `Priority Issue 9 to be Tackled
by Using the Post-K Computer' (Elucidation of The Fundamental Laws and Evolution of the Universe),
and Joint Institute for Computational Fundamental Science (JICFuS).
The computational resource on supercomputer Fugaku at RIKEN Center for Computational Science was provided through Usability Research ra000001.
The code development were partially performed on the supercomputer `Flow' at Information
Technology Center, Nagoya University.
The gauge configuration was provided through Japan Lattice Data Grid (JLDG)~\cite{JLDG}.

We dedicate this contribution to the memory of Yusuke Taniguchi,
who passed away in July 2022.
We praise his continuous effort as a core member of Bridge++ project since its launch
and being a driving force of the development.

\bibliographystyle{JHEP}
\bibliography{proc_lattice2022_bridge}

\end{document}